\documentclass[preprint,showpacs,preprintnumbers,amsmath,amssymb]{revtex4-2}
\usepackage{graphicx}
\usepackage{graphics}
\usepackage{epsfig}
\usepackage{verbatim}
\usepackage{amssymb}
\usepackage{xcolor}
\usepackage{gensymb}
\begin{document}
\title{Valley focusing effect in a rippled graphene superlattice}
\author{M. Pudlak}
\affiliation{Institute of Experimental
Physics,  04001 Kosice, Slovakia}
\author{R.G. Nazmitdinov}
\email{rashid@theor.jinr.ru}
\affiliation{Bogoliubov Laboratory of Theoretical Physics,
Joint Institute for Nuclear Research, 141980 Dubna,
Moscow region, Russia}
\affiliation{Dubna State University, 141982 Dubna, Moscow region, Russia}

\begin{abstract}
Graphene corrugations affect hybridization of $\pi$ and
$\sigma$ orbitals of carbon atoms in graphene based systems.
It can as well break differently the symmetry of the electron transfer integrals for different strip boundaries.
Using these facts, we found that the momentum distribution of electrons in ballistically propagating
beam can be selective without  external electric and/or magnetic fields in the graphene strip
under experimentally feasible periodic potential. Such a potential is created by means of the
superlattice that consists of periodically repeated graphene elements
(flat+rippled junction) with different
hybridization of carbon orbits, produced by variation of the graphene surface curvature.
As a result it gives rise to the valley dependent focusing effects
that can be controlled by alteration of number of superlattice elements.
\end{abstract}
\date{\today}
\maketitle

\section{Introduction}
\label{intro}
The exceptionally high charge carrier mobility in graphene has generated
enormous experimental and theoretical activity, with various potential applications
in nanotechnology in mind (see textbooks \cite{book1,book2}).
The remarkable graphene properties have been explained as
a consequence of linear energy dispersion of  the gapless low-energy excitations, provided
by graphene crystal structure that consists of two equivalent carbon sublattices.
Consequently, one can introduce graphene quasiparticles with different pseudospin quantum numbers
associated with the corresponding sublattices.
It is notable that this linear energy dispersion in  the low-energy spectrum of graphene is similar to the
Dirac-Weyl equation for massless neutrino \cite{An1}.

It was shown in \cite{ch1} that the
conservation of the pseudospin forbids strictly charged carrier
backscattering in a graphene monolayer with electrostatic
potential scattering that mimics the n–p junction. The barrier
always remains perfectly transparent for the normal incidence
of electrons, while the transmission decreases for other
angles. By virtue of this fact, electron focusing analogous to
optical effects that occur in negative refractive index material
is predicted \cite{ch2}. It is noteworthy to mention that above discussed results are
based on assumption of use external electrical or magnetic
accessories to control the focusing of electron flow.

We recall, however, that graphene sheets are not perfectly flat, and ripples
are considered as most natural sources that might be used
to control the electron mobility as well.
A number of proposals have been suggested in support of this idea.
It was predicted in the tight-binding approximation
that a corrugation (ripple)  could create in graphene electron scattering,
caused by the change in nearest-neighbor hopping parameters by the
curvature \cite{kg,gui}. Further on,
it was found that electrons in opposite valleys can be perfectly transmitted or totally reflected
in the presence of strain \cite{wu}. In Ref. \cite{jak}
it was shown how inhomogeneous strains can be used to create waveguides for valley polarized
transport of Dirac fermions in graphene.

Note, that the lattice deformation changes the distance between ions, $p_{z}$ orbital
orientation, and is leading to shift of the on-site energies of $p_{z}$
orbitals. This affects the effective Dirac equation that
could simulate the low energy electron states as a result of
a deformation-induced gauge field \cite{voz_pr}.
Moreover, the lattice deformation changes the relative
orientation of the orbitals
of the corrugated graphene sheet, leading to the hybridizations of
the $\pi$- and $\sigma$-bonds (see details in Appendix A).
The $\pi$ orbital dependence on the surface
curvature means that the local chemical potential varies with the
curvature. In fact, the hybridizations
leads to inhomogeneous charge distribution, and acts as
potential barriers for electrons leading to their
localization \cite{cm}.
This effect becomes important once it would be possible to create a
graphene system with controlled variation of the surface curvature.
In fact, the DFT and molecular dynamics simulations predict
that graphene sheet can be stretched up to about 20\%-30\%, without being
damaged \cite{kumar}.
The amplitude and the orientation of the unidirectional ripples
can be controlled with the aid of the applied strain \cite{jul1}.
And further, it was shown that using the hydrogenation it is possible
to induce periodic ripples with various thermal conductivity \cite{jul2}.

The discussed above theoretical ideas are supported, indeed, by
a few experimental techniques that demonstrate
evidently a spatial variation in graphene sheets nowadays.
The strain effect can be achieved by putting the material on
substrate that is micro-structured \cite{zhang1} or
mechanically deformed \cite{zhang2}.
Ripples can be formed by means of
the electrostatic manipulation
without any change of doping \cite{car19}.
Periodically rippled graphene can be fabricated by the epitaxial technique
(e.g., \cite{parga}). In this case,
in contrast to free-standing graphene,
a strong modification of the electronic structure of graphene is
observed, that gives rise to localized phonon \cite{phon} and
plasmon \cite{plaz} modes. Periodic nanoripples can be created
as well by means of the chemical vapor deposition \cite{ni}.
It is found that ripples or wrinkles act as potential barriers for
charged carriers leading to their localization \cite{car16}, in agreement with
the theoretical estimations \cite{cm}.

One of the main aims of the present paper is to demonstrate that strain effects
could provide the ability of the valleytronics to manipulate and detect the valley
degree of freedom of the ballistic electron transport.
To this aim we employ the model of rippled graphene superlattice
discussed in \cite{pe,sym}. In the present paper we extended this model by considering
the dependence of the hopping integrals between $\pi$ orbitals in zig-zag
and armchair graphene surface curvatures following the approach developed by Ando \cite{Ando}.
We recall that
a typical transition lengths for n-p junction are less 100 nm (e.g., \cite{huard}),
which allows to employ a ballistic transport model
for the study of physics n-p junction devices \cite{low}.
To demonstrate the effect of valleytronics various authors introduced
either electrical/magnetic fields or additional potentials to simulate
strain effects (see for a review \cite{chrev}).
We will show that the effective potential
determined by the variation of the local curvature of the graphene sheet
provides an additional design degree of freedom for both fundamental studies and
graphene-based electronic devices.

\section{Scattering model}
We suppose that incident ballistic electrons move from the left planar graphene
piece to the right planar piece passing through N elements of the superlattice.
 The unit element of the superlattice is composed of one ripple+one planar piece.
The graphene strip (the superlattice) is terminated in the x direction by zigzag or armchar boundaries,
while it infinitely long in the y direction.
Each element represents  a single junction described below.

\subsection{Eigenvalue problem}
\label{eigp}

\begin{figure}
\centerline{\includegraphics[width=0.95
\columnwidth]{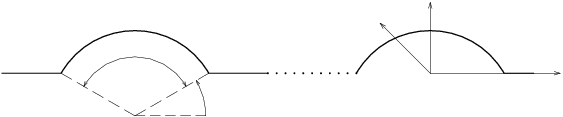}\unitlength=1mm
\begin{picture}(0,0)(0,0)
\put(-90,25.5){\makebox(0,0)[b]{$III$}}
\put(-120,25.5){\makebox(0,0)[b]{$II$}}
\put(-145,25.5){\makebox(0,0)[b]{$I$}}
\put(-110,0.8){\makebox(0,0)[b]{$\theta_0$}}
\put(-120,6.){\makebox(0,0)[b]{$\phi$}}
\put(-130,0.5){\makebox(0,0)[b]{$R$}}
\put(-2,8){\makebox(0,0)[b]{$x$}}
\put(-50,28){\makebox(0,0)[b]{$y$}}
\put(-35,34){\makebox(0,0)[b]{$z$}}
\end{picture}}
\caption{The corrugated graphene structure. The flat graphene pieces are located
in: the region (I), $-\infty<x< -R\cos\theta_{0}$,  $-\infty < y < \infty$;
and the region (III), $R\cos\theta_{0}<x<\infty$, $-\infty < y < \infty$.
 In the region (II), $-R\cos\theta_{0} <x< R\cos\theta_{0}$ and $-\infty < y < \infty$, we consider
a ripple (a curved surface in a form of arc of a circle).
The ripples are ordered in $x$ direction with the symmetry $y$-axis.}
\label{medium}
\end{figure}

The corrugated graphene structure is modelled by
a curved surface in a form of arc of a circle connected from
the left-hand and the right-hand sides to two flat graphene sheets (see Fig.\ref{medium}).
Hereafter, we consider a wide enough graphene sheet W$\gg$M, where W and M being, respectively,
as the width along the y axis and the length along x axis of the graphene sheet.
It means that we keep the translational invariance along the y axis and neglect the edge effects.

To analyse our junction, we take into
account: i) the variation of the hybridization of the carbon atom orbitals with  a surface curvature of a graphene sheet
(see Appendix A); ii)the modification of the electron transfer integrals, caused by the variation of the surface
curvature. The variation of the hybridization can be described by an effective electric potential
$\varepsilon(x)$.  The modification of the electron transfer integrals can be calculated as a shift of
a vector potential $\Delta\hat{k}$ in the matrix Hamiltonian in the effective
mass approximation. In order to find these modifications we extended the
approach, developed by Ando (see discussion in Ref.\cite{Ando}), and derived the corresponding
Hamiltonian (the details will be published elsewhere).
Thus, in the effective mass approximation the eigenvalue problem for
the envelope function can be written in the following form
\begin{eqnarray}
\label{ham}
&&\left(\begin{array}{cc}\varepsilon (x)&D_{\tau}\\
D_{-\tau}&\varepsilon
(x)\end{array}\right)
\left(\begin{array}{c}
F_{A}\\F_{B}\end{array}\right)=E
\left(\begin{array}{c}F_{A}\\F_{B}\end{array}\right)\,,
\\
&&D_{\tau}=\gamma[(\hat{k}_{x}+\Delta k_{x})-i\tau(\hat{k}_{y}+\Delta
k_{y})]\,,
\end{eqnarray}
where $\tau=+$ corresponds to $K$ point, while $\tau=-$ corresponds to for $K^{\prime}$ point.
The two components of the wavefunction refer to the two sublattices of carbon atoms.
The additional spin degeneracy of the excitations do not important in our consideration.

As it is shown in Appendix A,  the dependence of
energy $\varepsilon_{\pi}$ on the local surface curvature can be
expressed in the form
\begin{equation}
\label{elec} \varepsilon_{\pi} =\varepsilon_{2p}+ \alpha
\left(\frac{a}{R}\right)^{2}\,.
\end{equation}
Here $\varepsilon_{2p}=\langle p_{z}|H|p_{z} \rangle$ is the
$|p_{z}\rangle$ orbital energy of carbon atom, $R$ is the ripple
radius, and $\alpha=-0.58eV$.
Thus, the energy difference
between the $\pi$ orbitals in the curved and flat graphene is
\begin{equation}
\varepsilon=\varepsilon_{2p}-\varepsilon_{\pi}=\Delta \varepsilon =|\alpha |
\left(\frac{a}{R}\right)^{2}\approx
0.58\left(\frac{a}{R}\right)^{2}eV\,.
\label{endif}
\end{equation}
This difference can be considered as a contribution of the effective electric field produced
by the curvature dependence of the hybridization.
In the case of the flat graphene
$R\rightarrow\infty$ and, consequently, $\varepsilon (x)=\varepsilon_{2p}$.
The difference between $\varepsilon_{\pi}$ (curved region) and
$\varepsilon_{2p}$ (flat region) is important when the systems with
different surface curvature are coupled. Hereafter, for the sake of
simplicity we assume that $\varepsilon_{\pi}=0$.

In the flat region (I and III) the eigenstate of Eq. (\ref{ham})
has the following form
\begin{equation}
\label{4}
 F(x,y)=e^{ik_{x}x}e^{ik_{y}y}\frac{1}{\sqrt{2}}\left(\begin{array}{c}s
 e^{-i\tau\varphi}\\1\end{array}\right),\quad s=\pm 1\,,
\end{equation}
and the eigenenergy is
\begin{equation}
\label{5}
E=\varepsilon+s \gamma \sqrt{k_{x}^{2}+k_{y}^{2}}\,,
\end{equation}
where $s=+1(-1)$ is associated with the conductance (valence) band.
Here
\begin{equation}
\label{6}
e^{-i\varphi}=\frac{k_{x}-ik_{y}}{\sqrt{k_{x}^{2}+k_{y}^{2}}}\,.
\end{equation}
The eigenfunction and eigenenergy in the region II (a ripple) are obtained in the following forms
\begin{equation}
\label{7}
 F(x,y)=e^{i\kappa_{x}x}e^{ik_{y}y}\frac{1}{\sqrt{2}}\left(\begin{array}{c}s e^{-i\tau\chi}\\1\end{array}\right)\,,
\end{equation}
\begin{equation}
\label{8}
E= s \gamma \sqrt{(\kappa_{x}+\Delta
k_{x})^{2}+(k_{y}+\Delta k_{y})^{2}}\,.
\end{equation}
Here
\begin{equation}
\label{9}
 e^{-i\chi}=\frac{(\kappa_{x}+\Delta
k_{x})-i(k_{y}+\Delta k_{y})}{\sqrt{(\kappa_{x}+\Delta
k_{x})^{2}+(k_{y}+\Delta k_{y})^{2}}}\,.
\end{equation}
The vector field $\Delta \vec{k}$ depends on the graphene surface
curvature. The concrete form of this field will be used to
investigate the transport properties of a corrugated graphene with
zig-zag and armchair boundaries.

\begin{figure}
\includegraphics[width=0.8\textwidth]{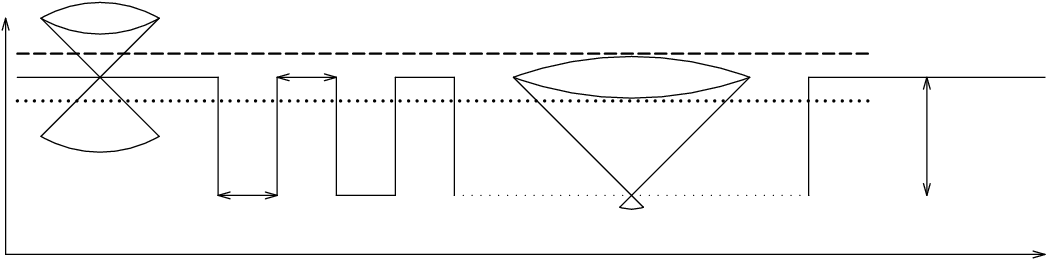}
\unitlength=1mm\begin{picture}(0,0)(0,0)
\put(-15,20){\makebox(0,0)[b]{$\varepsilon$}}
\put(-93,26){\makebox(0,0)[b]{$L_1$}}
\put(-102,2){\makebox(0,0)[b]{$L_2$}}
\put(1,1){\makebox(0,0)[b]{$x$}}
\put(-132,34){\makebox(0,0)[b]{$E$}}
\put(-22,26){\makebox(0,0)[b]{$B$}}
\put(-23,19){\makebox(0,0)[b]{$A$}}
\end{picture}
\caption{A schematic illustration of the scattering process in the superlattice
that composed of $N$ elements. Each element contains the flat region of
the length $L_{1}$ connected to the ripple of the  length $L_{2}$.
The energy $\varepsilon$ is brought about by the curvature dependence of
the hybridization effect (see text).
For a separate undoped flat graphene sheet $(\cal{F})$,
the Fermi energy lies exactly at the Dirac point $\varepsilon_{2p}$ (indicated by
a solid line that crosses the  Dirac point, the left side).
Similar picture takes place for a separate curved graphene piece $(\cal{C})$, where the
position of the Fermi energy $\varepsilon_{\pi}$, indicated by the dotted line on the right
side. For the hybrid system that consists
of $\cal{F}+\cal{C}$ pieces, there are two cases with the electron incident energies indicated by the lines A and B.}
\label{N}
\end{figure}

We assume that the flat segment has the length $L_{1}$, while the
a ripple has the length $L_{2}=R\phi$, see
Figs.\ref{medium},\ref{N}.
We consider the scattering at the interface introduced by different hybridizations
in the flat and the curved graphene regions. The interface is assumed to be smooth
on the length scale of a graphene unit cell (an inverse Brilloin momentum $2\pi/K$).
Consequently, it does not induce the intervalley ($K\rightarrow K^{\prime})$
scattering.

\subsection{Transport phenomena}

\subsubsection{Single junction}

Before to analyse the ballistic transport through the superlattice ${\cal{S}}$ we have to consider
transmission of electron, travelling with energy $E$ through
the hybrid subsystem $\cal{F}+\cal{C}$ (the unit element), at two most typical cases: above and below
the effective potential $\varepsilon$ (see Fig.\ref{N}).
In the both cases we have to match the corresponding wave functions in the flat and curved graphene pieces.

In the first flat ($\cal{F}$) sector of the ${\cal{S}}$ region
$[X_L \leq x <X_L+L_{1}$, $|y|<W]$ we consider the wave function in the form
\begin{equation}
\label{11}
\footnotesize \Psi
(x,y)=\frac{e^{ik_{y}y}}{\sqrt{2}}\left\{e^{ik_{x}x}\left(\begin{array}{c}e^{-i\varphi}\\1\end{array}\right)
+r e^{-ik_{x}x}\left(\begin{array}{c}-e^{i\varphi}\\1\end{array}
\right)\right\}\,,
\end{equation}
and for the first rippled ($\cal{C})$ sector of the ${\cal{S}}$ region  $[X_L+L_{1} \leq x < X_L+L_{1}+L_{2}$,
$|y|<W]$ we define the wave function in the form
\begin{equation}
\label{12} \footnotesize \Psi
(x,y)=\frac{e^{ik_{y}y}}{\sqrt{2}}\left\{\alpha_{1}e^{i\kappa_{x}x}\left(\begin{array}{c}e^{-i\chi}\\1\end{array}\right)+
\beta_{1}e^{-i\kappa_{x}x}\left(\begin{array}{c}-e^{i\chi}\\1\end{array}\right)\right\}\,,
\end{equation}
and so on. For the last flat region $[X_L+N(L_1+L_2)\leq x < M$, $|y|<W]$ we have
\begin{equation}
\label{14} \Psi (x,y)=
\frac{e^{ik_{y}y}}{\sqrt{2}}te^{ik_{x}x}\left(\begin{array}{c}e^{-i\varphi}\\1\end{array}
\right)\,.
\end{equation}
The unknown coefficients
$\alpha_{i},\beta_{i}$ can be obtained from the
continuity conditions on the boundaries.

Let us consider the specific features of the transmission, related to the incident electron energy $E$
with regard to the effective potential $\varepsilon$.
\begin{itemize}
\item
$E>\varepsilon$.

In this case we have the following condition for the incident electron energy $E$
(denoted as a line B in Fig.\ref{N}):
\begin{equation}
\label{10}
E= \varepsilon+ \gamma \sqrt{k_{x}^{2}+k_{y}^{2}}=\gamma
\sqrt{(\kappa_{x}+\Delta k_{x})^{2}+(k_{y}+\Delta
k_{y})^{2}}\,.
\end{equation}
For this particular case we obtain for transmission coefficients
\begin{eqnarray}
\label{20}
&T_{11}=e^{ik_{x}L_{1}}\left[\cos\kappa_{x}L_{2}-
i\sin\kappa_{x}L_{2}f^{(-)}\right]\;\,,\\
\label{st11}
&T_{11}=T_{22}^{*}\;\,,
\end{eqnarray}
where
\begin{equation}
\label{fpm}
f^{(\pm)}=\frac{1\pm\sin\varphi\sin\chi}{\cos\varphi\cos\chi}\;\,.
\end{equation}
And
\begin{eqnarray}
\label{21}
T_{21}&=&e^{-i\varphi}e^{ik_{x}L_{1}}\sin\kappa_{x}L_{2}\frac{\sin\chi
-\sin\varphi}{\cos\varphi\cos\chi}\;\,,\\
\label{st21}
T_{21}&=&T_{12}^{*}\,.
\end{eqnarray}
 Based on the above relations,
we have the following useful equation
\begin{eqnarray}
\label{22}
&&\frac{T_{11}+T_{22}}{2}=a_1-a_2f^{(-)}\;\,,\\
\label{a1}
&&a_1=\cos k_{x}L_{1}\cos\kappa_{x}L_{2}\;\,,\\
\label{a2}
&&a_2=\sin k_{x}L_{1}\sin\kappa_{x}L_{2}\,.
\end{eqnarray}

The valley dependence relations between quantities $k_{y},k_{x},\kappa_{x}$ are
defined by  Eq.(\ref{10}).

\item
$0< E <\varepsilon$.

In this case we have the following condition for the
incident electron energy $E$ (denoted as a line
A in Fig.\ref{N}):
\begin{equation}
\label{25} E= \varepsilon - \gamma \sqrt{k_{x}^{2}+k_{y}^{2}}=\gamma
\sqrt{(\kappa_{x}+\Delta k_{x})^{2}+(k_{y}+\Delta
k_{y})^{2}}\,.
\end{equation}
For transmission coefficients we obtain the following results
\begin{eqnarray}
\label{26}
&&T_{11}=e^{-ik_{x}L_{1}}\left[\cos\kappa_{x}L_{2}+i\sin\kappa_{x}L_{2}f^{(+)}\right]\;\,,\\
\label{27}
&&T_{21}=\;\;e^{-i\varphi}e^{ik_{x}L_{1}}\sin\kappa_{x}L_{2}\frac{\sin\chi
+\sin\varphi}{\cos\varphi\cos\chi}\,.
\end{eqnarray}
The same relations Eqs.(\ref{st11}), (\ref{st21}), are valid as well, that
yield another useful equation
\begin{equation}
\label{28}
\frac{T_{11}+T_{22}}{2}=a_1+a_2f^{(+)}\,.
\end{equation}

\end{itemize}
Here, the relations between $k_{y},k_{x},\kappa_{x}$ are defined by
Eq.(\ref{25}).

\subsubsection{Superlattice}

Using the continuity conditions on boundaries, we arrive to the equations
for the transmission coefficient $t$ through the
block of $N$ ripples and the corresponding reflection coefficient
$r$
\begin{equation}
\label{15}
\left(\begin{array}{c}1\\r\end{array}\right)=\left(\begin{array}{cc}T_{11}&
T_{12}\\T_{21}&
T_{22}\end{array}\right)^{N}=\left(\begin{array}{cc}N_{11}&
N_{12}\\N_{21}&
N_{22}\end{array}\right)\left(\begin{array}{c}t\\0\end{array}\right)\,.
\end{equation}
These equations yield the obvious relations
\begin{equation}
\label{16} t=1/N_{11}; \ \ r=N_{21} \ t \,.
\end{equation}

With the aid of Eqs.(\ref{20})-(\ref{st21}), (\ref{26}), (\ref{27}) it is readily to show that
\begin{equation}
\label{unit}
\det{(T)}=1\,.
\end{equation}
We recall that the $T$-matrix is subject to the condition
\begin{equation}
\label{cont}
\bigg(
\begin{array}{cc}
T_{11} & T_{12}
\\
T_{21} & T_{22}
\end{array}
\bigg)\bigg(
\begin{array}{cc}
a \\ b
\end{array}
\bigg)= \lambda \bigg(
\begin{array}{cc}
a \\ b
\end{array}
\bigg)\,.
\end{equation}
With the aid of Eqs.(\ref{unit}), (\ref{cont}), we obtain
the eigenvalues
\begin{equation}
\label{l12}
\lambda_{1,2}=\beta\pm\sqrt{\beta^2-1}\,,\quad \beta=(T_{11}+T_{22})/2\,.
\end{equation}

The transformation $U$ (that  diagonalizes the matrix $T$)
\begin{equation}
\label{ut}
U=\left(\begin{array}
{cc} a_1& a_2\\
b_1 & b_2
\end{array}\right)\Rightarrow
U^{-1}TU=
\left(\begin{array}{cc}\lambda_1&0\\0&\lambda_{2}
\end{array}\right)
\end{equation}
yields, in virtue of Eqs.(\ref{15}), (\ref{ut}), the following relation
\begin{equation}
U\left(\begin{array}{cc}\lambda_1^N&0\\0&\lambda_{2}^N
\end{array}\right)U^{-1}=
\left(\begin{array}{cc}N_{11}&
N_{12}\\N_{21}&
N_{22}\end{array}\right)
\,.
\end{equation}
By means of the standard procedure it is readily to obtain the matrices $U$ and $U^{-1}$ ($UU^{-1}=1)$.
As a result, taking into account that $\lambda_1\lambda_2=1$, we obtain the following definitions
\begin{eqnarray}
\label{n11}
&N_{11}=\frac{T_{11}(\lambda_{1}^{N}-\lambda_{2}^{N})+\lambda_{2}^{N-1}-\lambda_{1}^{N-1}}{\lambda_{1}-\lambda_{2}}=
N_{22}^{*}\;\,,\\
\label{n12}
&N_{12}=T_{12}\frac{\lambda_{2}^{N}-\lambda_{1}^{N}}{\lambda_{2}-\lambda_{1}}=N_{21}^{*}\;\,.
\end{eqnarray}
Evidently, the relation (\ref{15}) between the matrices $T$ and $N$,
and Eq.({\ref{unit}) yield the fulfilment of the following condition
\begin{equation}
\label{23}
\det{(N)}=\left|\begin{array}{cc}N_{11}& N_{12}\\N_{21}&
N_{22}\end{array}\right|=|N_{11}|^{2}-|N_{21}|^{2}=1\,.
\end{equation}
This secures that the condition $|r|^{2}+|t|^{2}=1$ is fulfilled,
 taking into account the definitions Eq.(\ref{16}).
As a result,
by means of Eqs.(\ref{16}), (\ref{23}), and the definition (\ref{n12}), we obtain the following expression
for the total transmission probability through $N$ elements of the superlattice
\begin{eqnarray}
\label{tr}
T_N&=&
|t|^{2}=\frac{1}{|N_{11}|^2}=\frac{1}{1+|N_{21}|^2}=\nonumber\\
&=&\frac{1}{1+|T_{12}|^{2}\left(\frac{\lambda_{2}^{N}-\lambda_{1}^{N}}{\lambda_{2}-\lambda_{1}}\right)^{2}}\,.
\end{eqnarray}

Evidently, the transmission probability (\ref{tr}) is a function of the incident electron energy
$E$ that determines the motion along the superlattice
(i.e., the wave numbers $k_x$ and $k_y$; see Eqs.(\ref{5}), (\ref{8})).
It is convenient to determine the transmission probability as a function of the wave number $k_y$
which together with the wave number $k_F=|E-\varepsilon|/\gamma$ determine details of electron transport.
With the aid of the transmission probability, the conductance is given by the Landauer formula
\begin{equation}
\label{trans}
G_{N}=4\frac{e^{2}}{h}\int_{-k_{F}}^{k_{F}}T_{N}(k_{y})\frac{dk_{y}}{2\pi/W}=4\frac{e^{2}}{h}\frac{k_{F}W}{\pi}I_{N}\,.
\end{equation}
Here, the integral $I_{N}$, defined by the expression
\begin{equation}
\label{73} I_{N}=\frac{1}{2}\int_{-1}^{1}T_{N}(u)du, \ \ u=\frac{k_{y}}{k_{F}}\,,
\end{equation}
characterizes the efficiency of the selection specific electron trajectories entering
into the considered system.

For the perfect transmission, i.e., for
$T(k_{y})=1$ the conductance
\begin{equation}
\label{72y}
G_{o}=4\frac{e^{2}}{h}\int_{-k_{F}}^{k_{F}}\frac{dk_{y}}{2\pi/W}=4\frac{e^{2}}{\pi
h}k_{F}W
\end{equation}
is the natural unit, since  $G_{N}=G_{o}I_{N}$.
For the discussion below we introduce the following terms
\begin{equation}
\label{gpm}
G_{N}^{+}=\frac{G_0}{2}\int_{0}^{1}T_{N}(u)du,\quad G_{N}^{-}=\frac{G_0}{2}\int_{-1}^{0}T_{N}(u)du\,.
\end{equation}

\section{Discussion}

Let us analyse  common and distinctive properties of the transport through the superlattice
with zig-zag and armchair graphene surface curvatures. In order to illuminate
the effect of the dependence on two interfaces,
we have to calculate the shift in the origin of $k_{x,y}$ by $\Delta k_{x,y}$,  produced by terms
of the order of $(a/R)^2$. In our analysis
we follow the arguments discussed by Ando (see $\S 5$ in \cite{Ando}).
The most interesting case is the transport phenomena at the incident electron energy
$0\leq E \leq \varepsilon$ (see Fig.\ref{N}, line A), which we study below in details.

\begin{itemize}
\item Zig-zag surface.

In the effective mass approximation
for the zig-zag interface we obtain
\begin{equation}
\Delta k_{x}=\mp \frac{a}{4\sqrt{3}R^{2}}\left(1-\frac{3}{8}\frac{\gamma'}{\gamma}\right)\,,
\end{equation}
\begin{equation}
\label{zky}
\Delta k_{y}=0\,,
\end{equation}
where the upper sign corresponds to the $K$ point, while the lower sign to the $K^{'}$ point.
The parameter $\gamma=\sqrt{3}\gamma_{0}a/2
=-\sqrt{3}V_{pp}^{\pi}a/2$, $\gamma^{'}=\sqrt{3}(V_{pp}^{\sigma}-V_{pp}^{\pi})a/2$, where
$V_{pp}^{\pi}$ and $V_{pp}^{\sigma}$ are the hopping integrals for
$\pi$ and $\sigma$ orbitals, $a$ is the length of the primitive
translation vector.
We recall that in our model it is assumed
that $V_{pp}^{\pi}\approx -3$ eV and $V_{pp}^{\sigma}\approx 5 $ eV.
 Therefore, we have $\gamma^{'}/\gamma \approx 8/3$, i.e., $\Delta k_{x}\approx 0$.
Thus, in the case of zig-zag interface the shifts are negligibly small, i.e.,
$\Delta k_{x}\approx 0, \quad \Delta k_{y}=0$. It seems that in this case
the symmetry between $K$ and $K^\prime$ is conserved.

\begin{figure}[h]
\centering
\includegraphics[height=6cm,clip=]{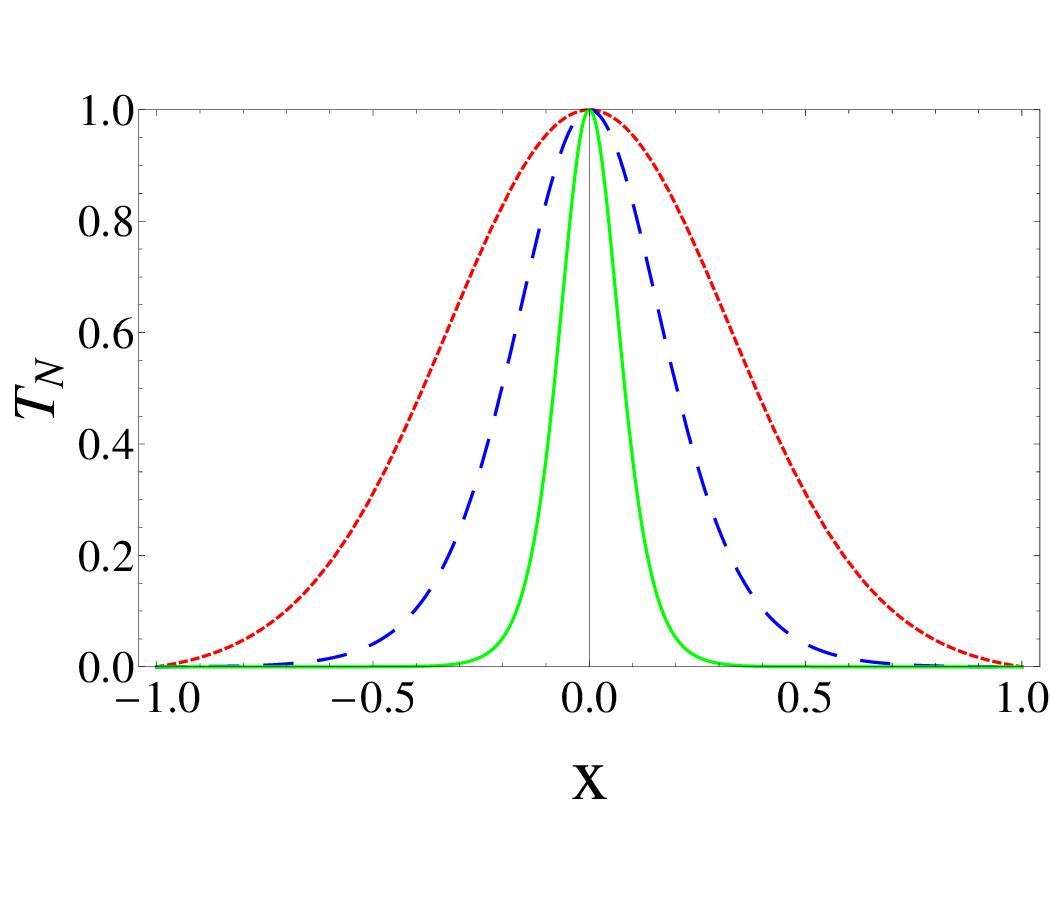}
\includegraphics[height=6cm,clip=]{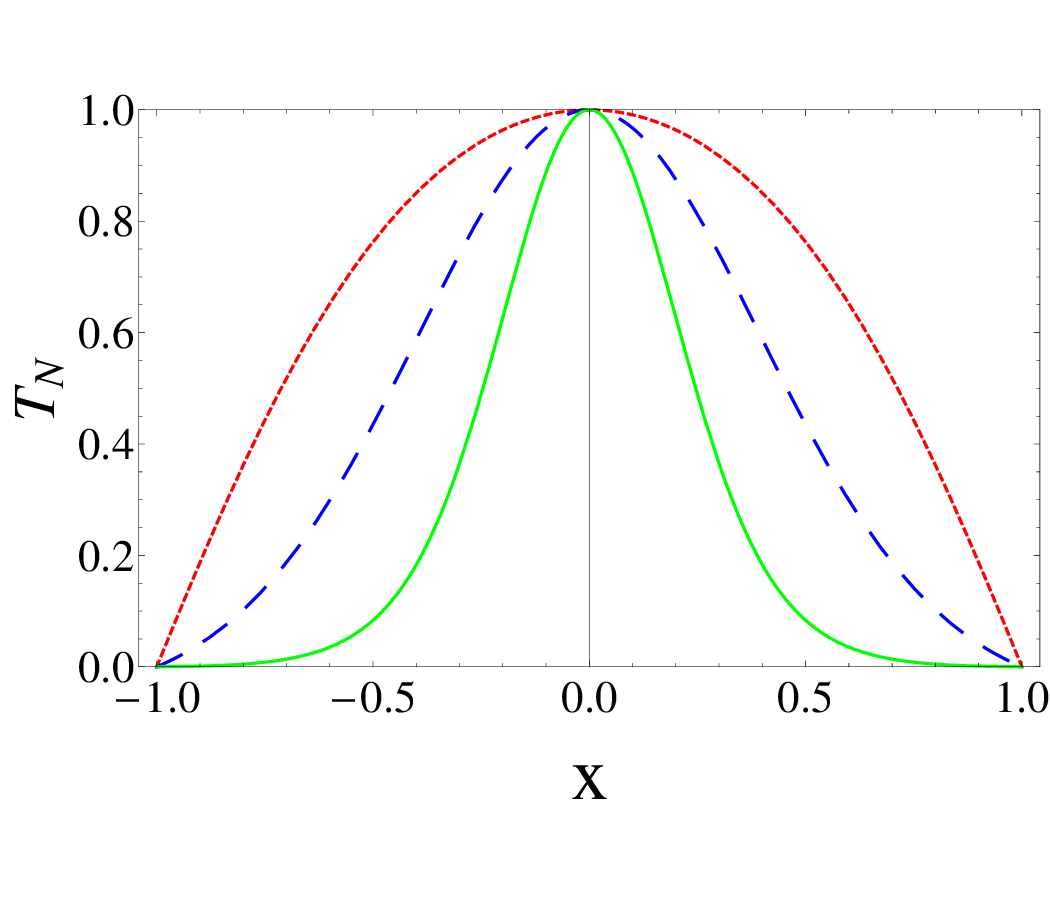}
\put(-235,123){\makebox(0,0)[b]{$(a)$}}
\put(-30,123){\makebox(0,0)[b]{$(b)$}}
\caption{Zig-zag surface. Transmission probabilities $T_N$ as
a function of $x=k_{y}/k_{F}$ for:
$N=10$ elements (dotted line);
$N=20$ elements (dashed line);
$N=50$ elements (solid line).
The energy of the incoming
electrons is $E=\varepsilon/2$, $L_{1}=10a$, $a\simeq 2.46$ \AA{}, $\phi=\pi$; (a)
$R=8$\AA{};\quad (b)$R=18$\AA{}. }
\label{fig4}
\end{figure}

In order to trace  the dependence of the transmission probability
on the incident angle of electrons, we calculate numerically Eq.(\ref{tr}) as a function
$k_{y}$ (see Fig.\ref{fig4}).
It is noteworthy that the superlattice leads to the selective transmission of electrons.
For a small number of N elements in the ${\cal S}$ subsystem the transmission probability
is nonzero for a wide range of values of $k_y$ (see results for $N=10,20$).
However, the larger the number of N elements in the superlattice, the stronger
the selectivity effect for ballistic electrons. Our system focuses the electronic flow,
selecting the transmission of those trajectories that are close to the normal incidence.
In fact, for a large enough number N elements of the superlattice
the selection does not depend on the incident direction of an electron flow at all !
Indeed, at $N\gg 1$, only for the direction perpendicular to the surface of the ${\cal S}$ subsystem
there is almost the ideal transmission,
while for the other angles ($k_y\neq 0$) there is only reflection.
Note, however, that the increase of the ripple radius decreases the selectivity effect (see Fig.\ref{fig4}b).
\begin{figure}[h]
\centering
\includegraphics[height=6cm,clip=]{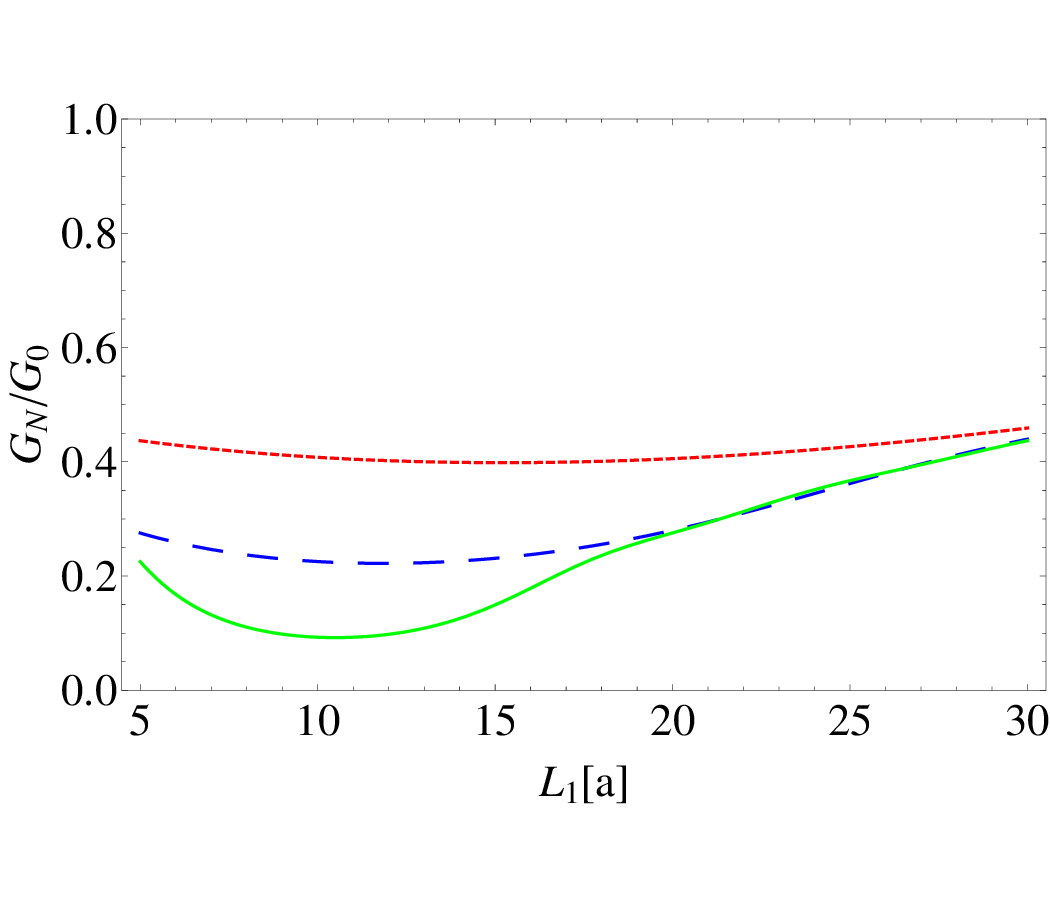}
\includegraphics[height=6cm,clip=]{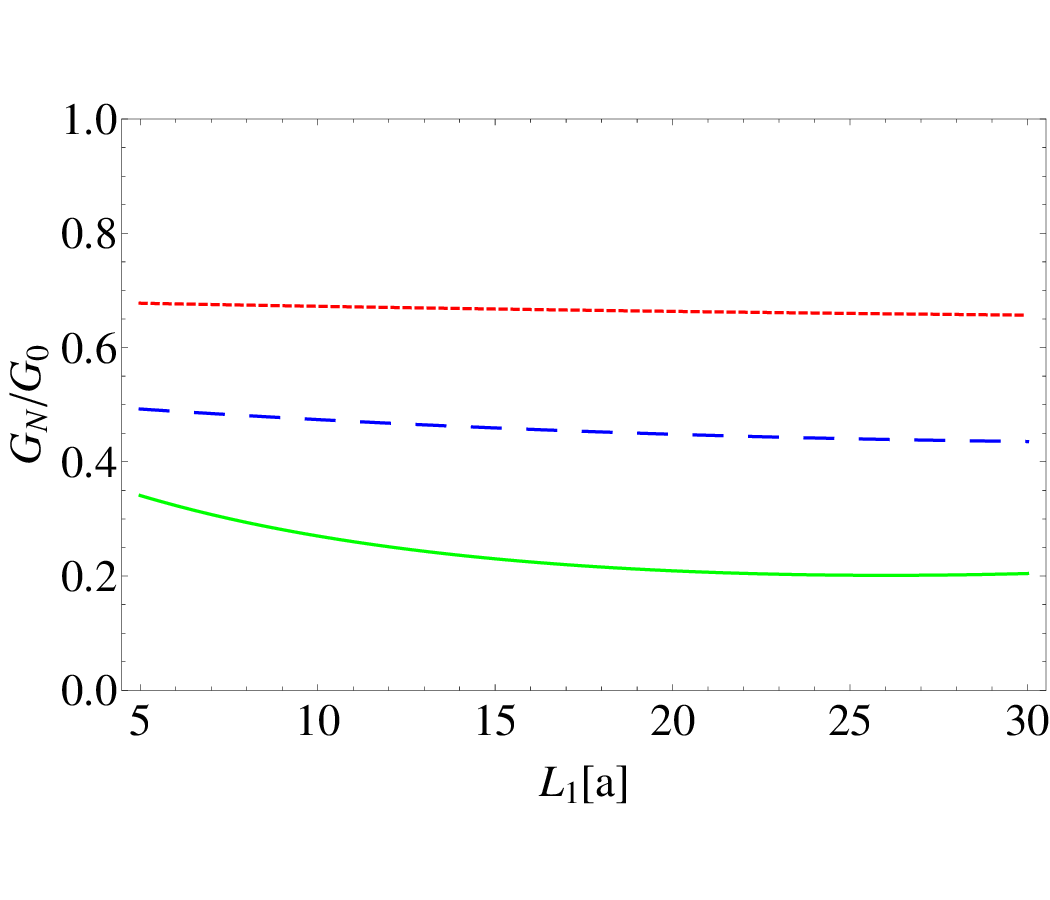}
\put(-235,123){\makebox(0,0)[b]{$(a)$}}
\put(-30,123){\makebox(0,0)[b]{$(b)$}}
\caption{The conductivity $G_{N}/G_{0}$ as a function of the length $L_{1}$ of the flat
region in $a$ units. Similar parameters are used as for Fig.\ref{fig4}.
}
\label{fig5}
\end{figure}

The selective electrons transmission across the interface
created by $N$ units is demonstrated on Fig.\ref{fig5}, where the
dependence of $G_{N}/G_{o}$ on the length of the flat region $L_1$
is depicted. The electron conductivity $G_{N}$ across the interface
with $N$ units is much smaller in comparison to $G_{o}$ for enough
large $N$ at relatively small value of the flat region $L_1\leq 10a$
in the superlattice at the small value of the ripple radius $R=8$ \AA{}.
With the increase of the flat region  $L_1\geq 18a$ the conductivity tends to
the limit manifested for small number of ripples, simultaneously
losing the selective properties. For a large ripple radius
$R=18$ \AA{} the selectivity effect and the conductivity decrease
with the increase of the flat region $L_1$ (see Fig.\ref{fig5}b).
Thus, there is an optimal set of parameters such as: the ripple radius $R$, the flat
region length $L_1$, the number of elements of the superlattice,
that provide the most efficient focusing effect. We return to this point below.

\item Armchair surface.

In the effective mass approximation
for the armchair interface we obtain
\begin{equation}
\label{41}
\Delta k_{x}=0\,,
\end{equation}
\begin{equation}
\label{42}
\Delta k_{y}={\mp}\frac{1}{4\sqrt{3}a}\left(\frac{a}{R}\right)^{2}\left(\frac{5}{8}\frac{\gamma'}{\gamma}-1\right)\sim
{\mp}\frac{1}{6\sqrt{3}a}\left(\frac{a}{R}\right)^{2}\,.
\end{equation}

\begin{figure}[h]
\centering
\includegraphics[height=8cm,clip=]{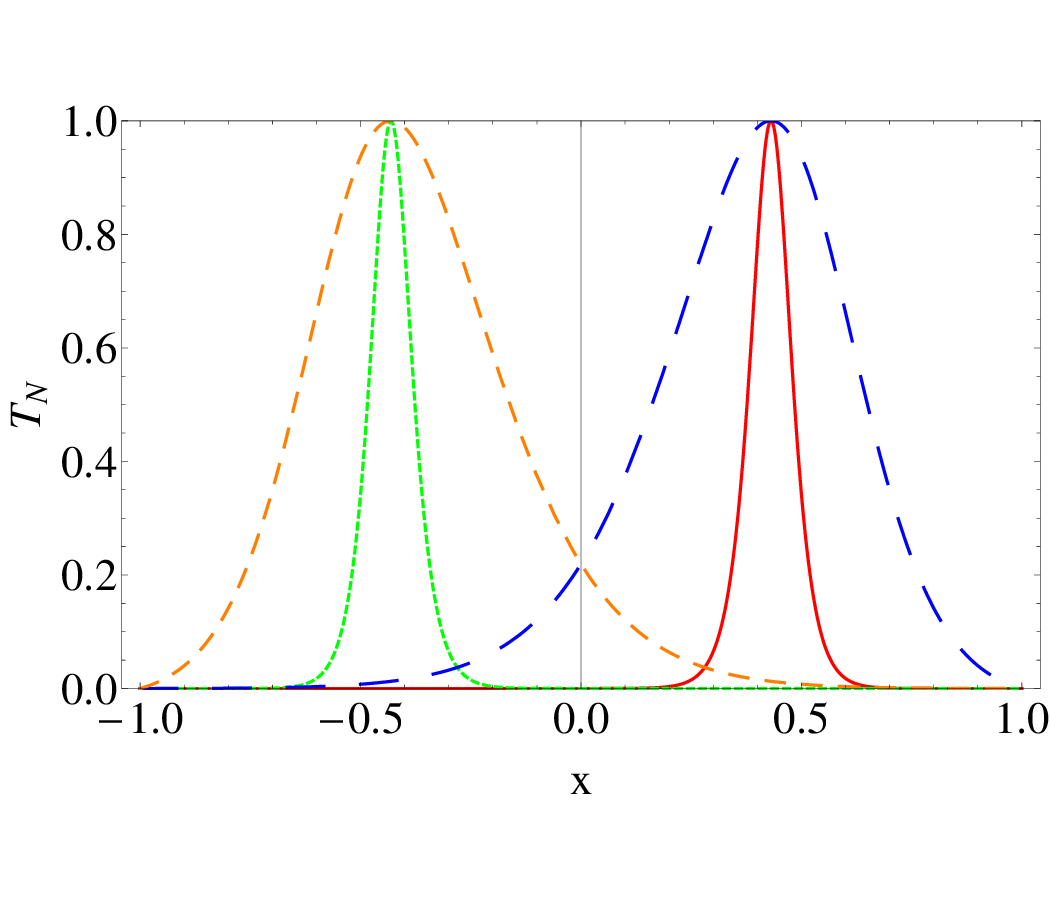}
\caption{Armchair surface. Transmission probabilities $T_N$ as
a function of $x=k_{y}/k_{F}$ for:
i)$N=15$ elements (valley $K$, long dashed line), (valley $K^\prime$, short dashed line);
ii)$N=70$ elements (valley $K$, solid line), (valley $K^\prime$, dotted line).
Filtering
in $\varphi$-space, induced by $N$ elements of the superlattice,
results in valley focusing effect. The action of a $\varphi$-filter allows
for electrons in valley $K$ to be transmitted freely,
while blocking them in valley $K^\prime$ at the same value of the angle $\varphi$.
The energy of the incoming
electrons is $E=\varepsilon/2$, $L_{1}=10a$, $a\simeq 2.46$ \AA{}, $\phi=\pi$,
$R=8$\AA{}.
}
\label{fig7}
\end{figure}

\begin{figure}[h]
\includegraphics[height=6cm,clip=]{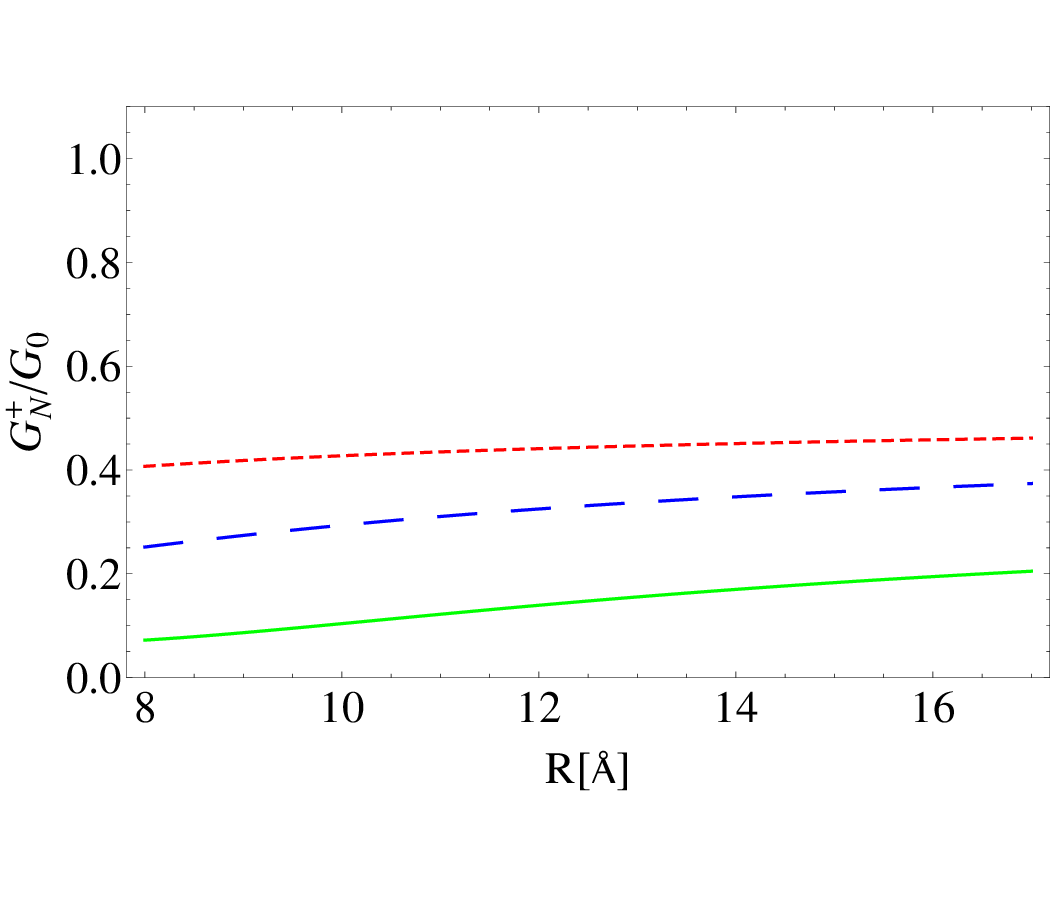}
\includegraphics[height=6cm,clip=]{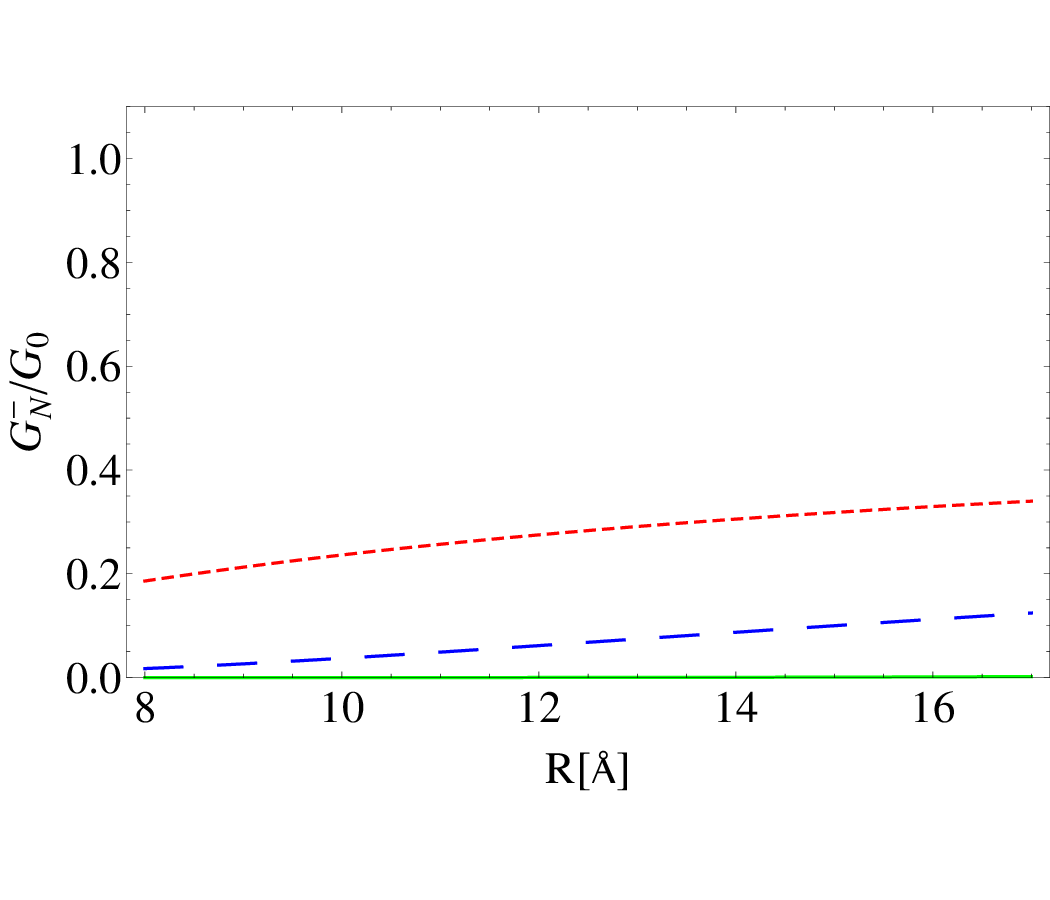}
\put(-235,123){\makebox(0,0)[b]{$(a)$}}
\put(-30,123){\makebox(0,0)[b]{$(b)$}}
\caption{
The conductivity $G_{N}^{\pm}/G_{0}$ as a function of
the ripple radius $R$ in $a$ units for: $N=5$ elements (dotted line);
$N=15$ elements (dashed line); $N=60$ (solid line).
The incident electron energy is
$E=\varepsilon/2$, $L_{1}=10a$, $\phi=\pi$, armchair surface.
The results are given for valley $K$.}
\label{Pol}
\end{figure}

In many ways, the transport properties of this system are similar to those of the zig-zag surface
(see Figs.\ref{fig4}, \ref{fig5}).
The basic difference consists in the asymmetry of the focusing effects in $K$ and $K^\prime$
valleys (see Fig.\ref{fig7}). It was speculated in Ref.\cite{fujita} that
the vector fields arising
from strain might be utilized to impose a valley-dependent filtering in a corrugated graphene sheet.
Indeed, our results  demonstrate evidently that the replacement  $\varphi\rightarrow -\varphi$
leads to a mirror image of the conductivity behavior in the other valley
($K\Leftrightarrow K^\prime$).
It is notable to mention that the electron conductivity in very similar to the one studied above.
Indeed, the conductivity decreases with increase of ripple radius (see Fig.\ref{Pol}). However,
for the zig-zag edge termination there is a  decrease of the both contribution $G_{N}^{\pm}/G_{0}$
around the position $x=0$. In contrast, with a total decrease of the conductivity in the system with
the armchair edge termination there is a prevailing of the
$G_{N}^{+}/G_{0}$ contribution over the $G_{N}^{-}/G_{0}$ contribution
 around the position of the supercollimation angle (see Figs.\ref{fig7},\ref{Pol})
with the increase of the number of $N$ elements. We return to this point below in detail.

\end{itemize}

To illuminate the basic features of the transport in the superlattice, let us compare
the selectivity effects of the latter case with that produced by the smooth step \cite{Allain,ch1}.
We recall that the estimation for the smooth step (which produces the focusing) yields the value
\begin{equation}
G_{sm}= 2\frac{e^{2}}{\pi
h}W\sqrt{\frac{k_{F}}{l}}=\frac{G_{o}}{2\sqrt{k_{F}\ell}}\,,
\end{equation}
that describes the  conductivity at the condition $k_{F}\ell\gg1$
($\ell$ is the step length).
In order to achieve
the smooth step effect, the corrugations with gradually increasing
curvature can be used in our case. This conditions leads to
the inequality
\begin{equation}
G_{sm} > G_N\Rightarrow \sqrt{2\pi\ell/\lambda_F}\times I_N<1/2\,.
\end{equation}
If we hold fixed the condition $\ell=N(L_1+L_2)$, this inequality
determines the number of elements  $N$ and their length $L_1+L_2$
at the same length $\ell$ for the
smooth potential and the superlattice. Thus, by appropriate choice
of the product $N(L_1+L_2)$ one can always use the
advantage  of electron flow focusing through  the superlattice, which number of elements can be
controlled externally. Moreover, one can use additionally the fine turning of the ripple radius
and change carrier charge densities on different sides of our hybrid system.

For completeness we present the results for transmission probabilities
at $E>\varepsilon$ for different surfaces (see Fig.\ref{fig9}).
Again, we observe the conservation of symmetry between $K$ and $K^\prime$ valleys in
a graphene sheet with a zig-zag surface, while it is broken in that with
an armchair surface. We found that the number of the ripples has a slight influence on the electron
transmission.

\begin{figure*}
\centering
\includegraphics[height=6cm,clip=]{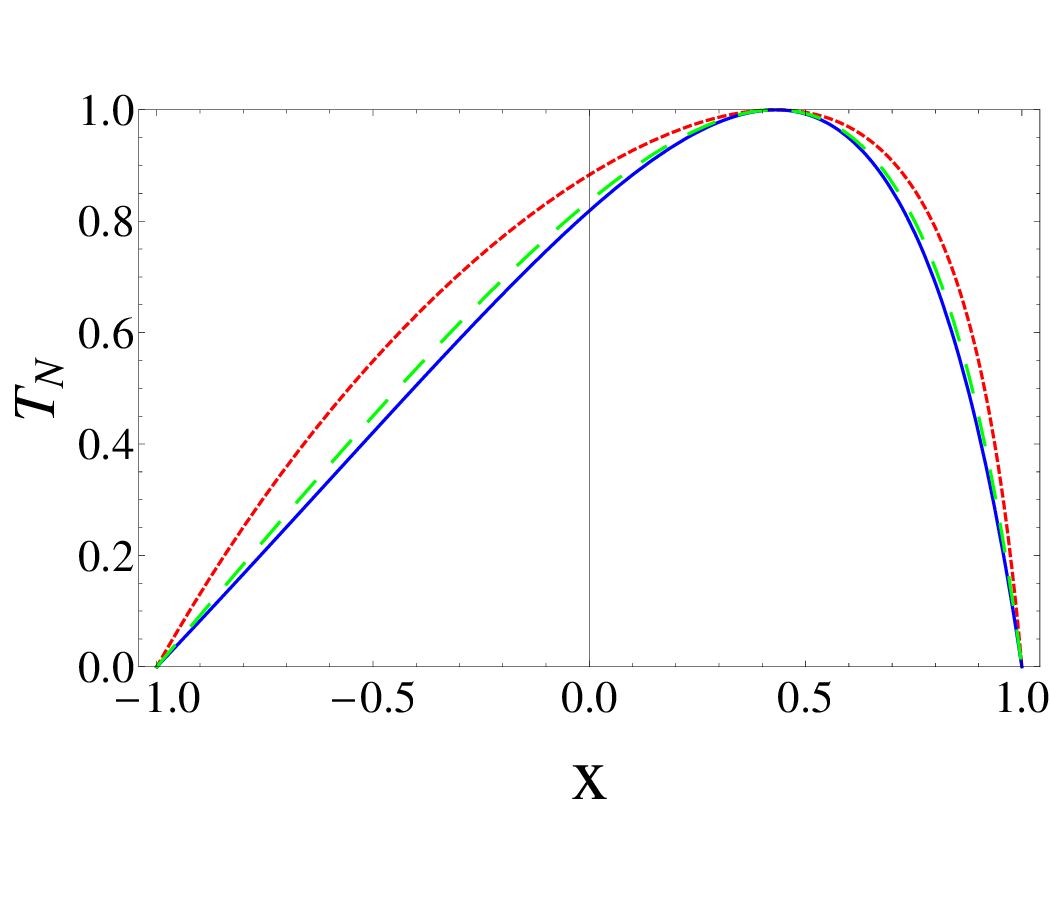}
\includegraphics[height=6cm,clip=]{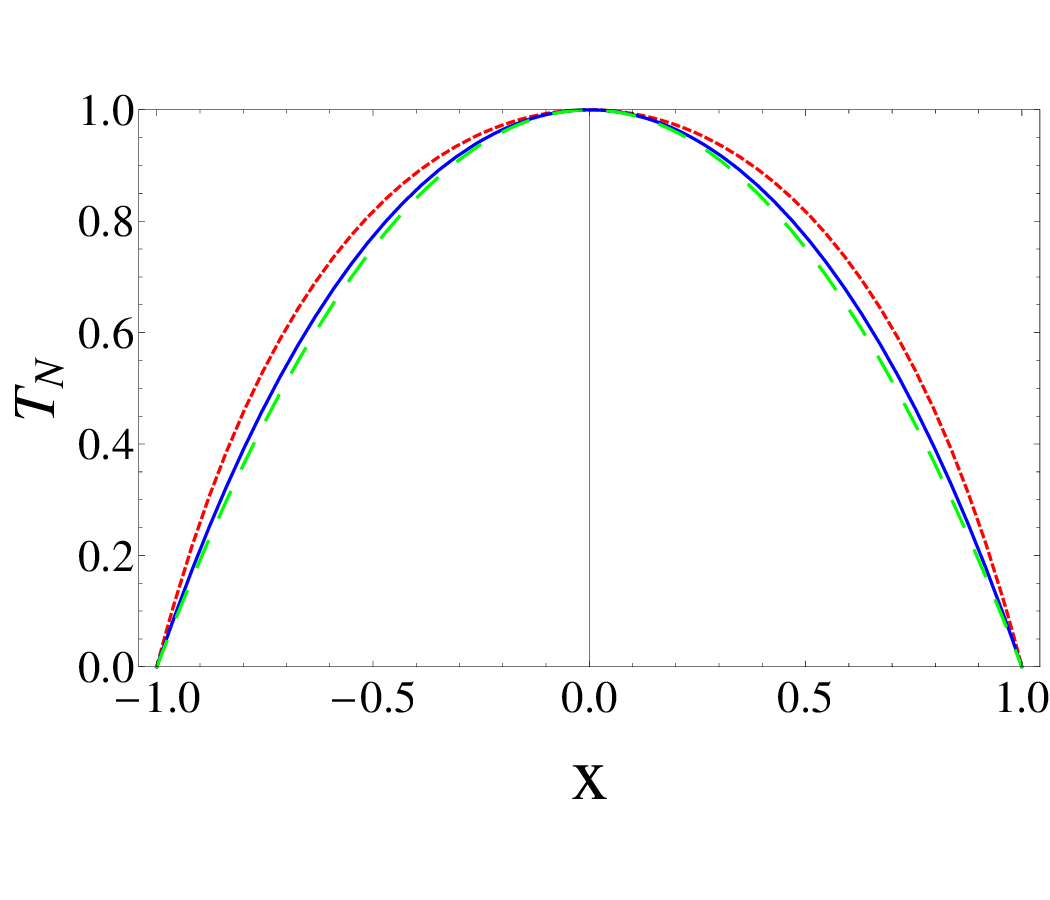}
\put(-235,123){\makebox(0,0)[b]{$(a)$}}
\put(-30,123){\makebox(0,0)[b]{$(b)$}}
\caption{Transmission probabilities $T_N$  as
a function of $x=k_{y}/k_{F}$ for $K^\prime$ valley. The energy of the incoming
electrons is $E=1.001\varepsilon$, $L_{1}=10a$, $R=18$\AA{},
$\phi=\pi$; (a) armchair surface; (b) zig-zag surface. The number of elements of the superlattice is
similar to Fig.\ref{fig4}.}
\label{fig9}
\end{figure*}

Depending on the  energy of the incident electron beam it is possible to determine analytically the angle
of the supercollimation in the case of the zig-zag and armchair edge terminations. Evidently, this angle is subject to the
condition $T_N=1\Rightarrow |T_{12}|^2\equiv|T_{21}|^2=0$ (see Eq.(\ref{tr})), which holds for the incident
electron energy $E>\varepsilon$ and  $E<\varepsilon$. Let us consider each case in detail.
\begin{itemize}
\item $E>\varepsilon$ :  In this case the condition $|T_{12}|^2=0 \Rightarrow \sin{\varphi}=\sin{\chi}$
(see Eq.(\ref{21})). Since the energy of incoming electron in the flat graphene piece
$E=\varepsilon+\gamma\sqrt{k_x^2+k_y^2}$, we have (see also Eq.(\ref{6}))
\begin{equation}
\sin{\varphi}=\frac{k_y}{(E-\varepsilon)/\gamma}\,.
\end{equation}
Taking into account the definition of  energy of transmitted electrons  in the curved graphene piece
$E=\gamma\sqrt{(\kappa_{x}+\Delta k_{x})^{2}+(k_{y}+\Delta k_y)}$, we have (see also Eq.(\ref{9}))
\begin{equation}
\label{schi}
\sin{\chi}=\frac{k_y+\Delta k_y}{E/\gamma}\,.
\end{equation}
As a result, we obtain
\begin{equation}
\sin{\chi}=\sin{\varphi} \Rightarrow x=\frac{k_y}{k_F}=\frac{\gamma \Delta k_y}{\varepsilon}
\end{equation}
where $k_F=(E-\varepsilon)\gamma$, and $\Delta k_y$ is determined by Eq.(\ref{42}).
\item $E<\varepsilon$.  In this case the condition $|T_{12}|^2=0 \Rightarrow  \sin{\varphi}=-\sin{\chi}$
(see Eq.(\ref{27})). Since in the flat graphene piece the energy of incoming electron (moving in the valence band,
see Fig.\ref{N})
is $E=\varepsilon-\gamma\sqrt{k_x^2+k_y^2}$ (see also Eq.(\ref{25})), we have
\begin{equation}
\sin{\varphi}=\frac{k_y}{(\varepsilon-E)/\gamma}\,.
\end{equation}
Taking into account Eq.(\ref{schi}), we obtain
\begin{equation}
\sin{\varphi}=-\sin{\chi} \Rightarrow x=\frac{k_y}{k_F}=-\frac{\gamma \Delta k_y}{\varepsilon}\,.
\end{equation}
\end{itemize}

Thus,  we obtain the definition of the supercollimation angle that is determined by the contribution
$\varepsilon$ produced by a curvature dependence hybridization, and by a magnitude of the vector
field $\Delta {\vec k}$ brought about by a graphene surface curvature.
In particular, at the incident electron energy $E<\varepsilon$  we obtain for valley $K$
\begin{equation}
\sin\varphi=-(\frac{\gamma \tau |\Delta k_y|}{\varepsilon})
=\left[\frac{\sqrt{3}}{2}a\gamma_{0}\frac{1}{6\sqrt{3}a}\left(\frac{a}{R}\right)^{2}\right]/\left[0.58\left(\frac{a}{R}\right)^{2}\right]\approx
0.43\,, \quad \tau=-1\,,
\end{equation}
which corresponds to the angle $\varphi\approx25.5^{\degree}$.

The existence of transmissions associated with valley quantum numbers raised the lovely discussion
on exploiting of the valley degree of freedom for  development carbon-based electronics named
graphene valleytronics  \cite{ry}.
Valley polarization, valley inversion \cite{23} and valley-contrasting
spatial confinement \cite{24} of massless Dirac fermions were
demonstrated experimentally in strained graphene under inhomogeneous
pseudomagnetic fields and tunable real magnetic fields.

The results for transmission at $N\gg 1$ and $\varepsilon > E$ (see Figs.\ref{fig4}, \ref{fig7})
indicate that propagating modes with the wavevector
$k\in[0,k_F]$ lie in the $K$ valley, whereas modes with the wavevector $k\in[0,-k_F]$
lie in the $K^\prime$ valley. At $\varepsilon < E$ we have the opposite situation.
Therefore, for the sake of discussion we consider the most interesting case $\varepsilon > E$.
With the aid of Eqs.(\ref{tr}, \ref{trans}, \ref{72y}) the valley polarization of the transmitted current
is quantified by
\begin{equation}
\label{pol}
P_{N}^{\tau}=\frac{\int_{0}^{k_{F}}T_{N}^{\tau}(k_{y},\tau \xi)\frac{dk_{y}}{2\pi/W}-
\int_{-k_{F}}^{0}T_{N}^{\tau}(k_{y},\tau \xi)\frac{dk_y}{2\pi/W}}
{\int_{-k_{F}}^{k_{F}}T_{N}^{\tau}(k_y),\tau \xi)\frac{dk_{y}}{2\pi/W}}
\end{equation}
where $\xi =|\Delta k_{y}|$, and the valley dependent $\Delta k_y$  is defined by Eqs.(\ref{zky}, \ref{42});
$\tau=(-/+)$ corresponds to $K/K^{\prime}$ valley, respectively. Taking into account the results of
calculations for the transmission probability (see Fig.\ref{fig4}), we obtain that $P_N^{K}=P_N^{K^{\prime}}=0$
for the superlattice with zig-zag edge termination for the both valleys. It is notable, that the conductivity is decreasing with
the increase of the number $N$-elements, focusing between $\varphi\leq |20^{\degree}|$, being symmetric for
the superlattice with zig-zag edge termination.
For superlattice with armchair edge termination
the polarization $P\in[-1,1]$, with $P=1$ if the transmitted current lies fully in the $K$ valley
and $P=-1$ if it lies fully in the $K^\prime$ valley (see Fig.\ref{fig7}).
In this case the conductivity decreases with the increase of the number of $N$-elements of the superlattice.

\section{Summary}

Based on the fact of the different type of hybridization of carbon atom orbitals in
the flat and the corrugated graphene pieces, we developed the  model that simulate
 n-p junction by means of the superlattice and describes the valley focusing effect.
In the approximation of the effective mass Hamiltonian,
the curvature dependence of the $\pi$-orbitals yields the variation of
the local chemical potential. This fact corresponds to the
effective electric field that depends on the electron localization.
The variation of the graphene curvature affects the transfer integrals as well,
that together with the hybridization provides the necessary conditions for
the implementation of the valley focusing effect.
It is notable that the modification of the transfer integrals becomes important
in the corrugated graphene sheet with the armchair surface, while it negligible
in the case of corrugated sheet with the zig-zag surface.

Our analysis of the superlattice system that consists of the periodically
repeated flat+ripple pieces demonstrates the strong selectivity
effect of transmitted electron trajectories with the increase of number $N$
(elements of the superlattice).
This effect becomes essential for incident electrons, moving in the energy interval
$0< E < \varepsilon$; where $\varepsilon$ is the energy difference
between the $\pi$ orbitals in the curved and flat graphene sheet.
The ballistic electron transmission depends on the radius of the
ripple, on the length of the arc of the ripple and on the width of
the flat region between ripples.
It is remarkable, however,
that in a multi-rippled graphene structure the maximum of the transmission
for the both valleys is reached at different angles that are characteristic constants:
$\varphi=0^{\degree}$ for the structure with zig-zag edge termination;
$|\varphi|\approx 25.5^{\degree}$ for the one with armchair termination.
The superlattice, described in the
paper, enables to one to control the filtering effect without
\textit{any additional electrical or magnetic sources}.
The larger is the number of elements N, the stronger is the selectivity.
At $N\gg 1$, only for the direction perpendicular to the surface of the ${\cal S}$ subsystem
there is almost the ideal transmission,
while for the other angles ($k_y\neq 0$) there is the strong reflection
for the superlattice with zig-zag edge termination.
In the superlattice with armchair edge termination similar filtering
takes place at the supercollimation angle $|\varphi|\approx 25.5^{\degree}$.
This phenomenon is due to the Klein tunneling that is grown in our system by virtue of
controlled graphene surface curvature.

\section*{Acknowledgments}
Michal Pudlak acknowledges the financial support by Slovak Grant Agency for Science VEGA under
the grant number VEGA 2/0076/23.

\appendix

\section{Hybridization in a curved graphene}
\label{ap1}

Let us compare the hybridization of $\pi$ and $\sigma$ orbitals
in the flat and curved graphene systems.
We consider the Hamiltonian
for the $K$ point  (similar approach can be applied for
$K^{'}$ point). It depends on two operators
$\hat{k}_{x}=-i\frac{\partial}{\partial x}$,
$\hat{k}_{y}=-i\frac{\partial}{\partial y}$, and
yields the equation for the envelope function of the flat graphene
in the effective mass approximation (e.g., \cite{SDD})
\begin{equation}
\label{1} \left(\begin{array}{cc}\varepsilon_{2p}&\gamma
(\hat{k}_{x}-i\hat{k}_{y})\\\gamma
(\hat{k}_{x}+i\hat{k}_{y})&\varepsilon_{2p}
\end{array}\right)\left(\begin{array}{c}
F^{K}_{A}\\F^{K}_{B}\end{array}\right)=E
\left(\begin{array}{c}F^{K}_{A}\\F^{K}_{B}\end{array}\right)\,.
\end{equation}
Here, the parameter $\gamma =\sqrt{3}\gamma_{0}a/2$ depends on  the length of the
primitive translation vector $a=\sqrt{3}d\simeq 2.46$ \AA{} with
$d$ being the distance between atoms in the unit cell, and it is
assumed that $\gamma_{0}\approx 3$ eV. The energy
$\varepsilon_{2p}=\langle 2p_{z}|\verb"H"|2p_{z} \rangle$
is the energy of $2p_{z}$-orbitals of carbon atoms in the flat graphene,
directed perpendicular to the graphene surface;
$\verb"H"$ is the tight-binding Hamiltonian of the graphene.
The solution of Eq. (\ref{1}) determines the wave function
\begin{eqnarray}
\label{wff}
&&F(x,y)=e^{ik_{x}x}e^{ik_{y}y}\frac{1}{\sqrt{2}}\left(\begin{array}{c}s
e^{-i\varphi}\\1\end{array}\right),\\
&&e^{-i\varphi}=(k_{x}-ik_{y})/\sqrt{k_{x}^{2}+k_{y}^{2}}
\nonumber
\,,
\end{eqnarray}
and the energy
\begin{equation}
E=\varepsilon_{2p}+s \gamma \sqrt{k_{x}^{2}+k_{y}^{2}}\,.
\end{equation}
Here, the sign $s=-1(+1)$ is associated with the valence (conductance) band.
In the flat graphene we have the following hybridization of $\pi$ and
$\sigma$ orbitals:
\begin{eqnarray}
&|\pi\rangle = |2p_{z}\rangle\,,\\
&|\sigma_{1}\rangle = \frac{1}{\sqrt{3}}|2s\rangle
+\sqrt{\frac{2}{3}}|2p_{y}\rangle\,,\\
&|\sigma_{2}\rangle = \frac{1}{\sqrt{3}}|2s\rangle +
\sqrt{\frac{2}{3}}\left(\frac{\sqrt{3}}{2}|2p_{x}\rangle
-\frac{1}{2}|2p_{y}\rangle \right)\,,\\
&|\sigma_{3}\rangle = \frac{1}{\sqrt{3}}|2s\rangle -
\sqrt{\frac{2}{3}}\left(\frac{\sqrt{3}}{2}|2p_{x}\rangle
+\frac{1}{2}|2p_{y}\rangle \right)\,.
\end{eqnarray}

Let us discuss in details the hybridization of
$\sigma$ and $\pi$ orbitals in the graphene with nonzero curvature.
The $\sigma$ orbitals create the bonds between carbon atoms, while the
$\pi$ orbitals determine the electronic properties of the graphene.
For the sake of illustration we  consider a zig-zag nanotube.

For the curved graphene (the arc, characterised by the
radius $R$) we obtain the space coordinates of the three
nearest-neighbor vectors $\vec{\tau}_{i}$ in the following form:
\begin{eqnarray}
&\vec{\tau}_{1}=d(0,1,0) \,,\\
&\vec{\tau}_{2}=d(\frac{\sqrt{3}}{2}\cos\vartheta
,-\frac{1}{2},-\frac{\sqrt{3}}{2}\sin\vartheta )\,,\\
&\vec{\tau}_{3}=d(-\frac{\sqrt{3}}{2}\cos\vartheta
,-\frac{1}{2},-\frac{\sqrt{3}}{2}\sin\vartheta )\,,
\end{eqnarray}
where $\sin\vartheta =a/4R$. At the limit $R\rightarrow \infty$,  the vectors
$\vec{\tau}_{i}$ transform to those of the flat graphene. Evidently, the
$\sigma_{i}$ -orbitals are determined by the vectors $\vec{\tau}_{i}$.
As a result, the $\sigma_{i}$ and $\pi$ orbitals can be expressed
as follows
\begin{eqnarray}
&|\pi\rangle = d_{1}|2s\rangle +d_{2}|2p_{x}\rangle
+d_{3}|2p_{y}\rangle +d_{4}|2p_{z}\rangle\,,\\
&|\sigma_{1}\rangle = c_{1}|2s\rangle
+\sqrt{1-c_{1}^{2}}|2p_{y}\rangle\,,\\
&|\sigma_{2}\rangle = c_{2}|2s\rangle +
\sqrt{1-c_{2}^{2}}\left(|\chi_1\rangle
-|\chi_2\rangle\right)\,,\\
&|\sigma_{3}\rangle = c_{3}|2s\rangle -
\sqrt{1-c_{3}^{2}}\left(|\chi_1\rangle
+|\chi_2\rangle\right)\,,\\
&|\chi_1\rangle=\frac{\sqrt{3}}{2}\cos\vartheta |2p_{x}\rangle\,,\\
&|\chi_2\rangle
= \frac{1}{2}|2p_{y}\rangle-\frac{\sqrt{3}}{2}\sin\vartheta
|2p_{z}\rangle \,.
\end{eqnarray}

With the aid of the orthonormality
conditions $\langle\sigma_{i}|\sigma_{j}\rangle =\delta_{ij}$,
$\langle \pi|\sigma_{j}\rangle =0$, and $\langle \pi|\pi\rangle =1$,
we determine the parameters $\{c_{k},d_{l}\}$ and obtain
the following expressions for the $\pi$ and $\sigma$ orbitals in
the lowest order of the ratio $a/R$:
\begin{eqnarray}
\label{1f}
&|\pi\rangle \approx |2p_{z}\rangle +\frac{a}{2\sqrt{6}R}|2s\rangle
+\frac{a}{4\sqrt{3}R}|2p_{y}\rangle\,,\\
\label{2f}
&|\sigma_{1}\rangle = \frac{1}{\sqrt{3}}|2s\rangle
+\sqrt{\frac{2}{3}}|2p_{y}\rangle\,,\\
\label{3f}
&|\sigma_{2}\rangle = \frac{1}{\sqrt{3}}|2s\rangle +
\sqrt{\frac{2}{3}}\left(\frac{\sqrt{3}}{2}|2p_{x}\rangle -
|\chi_{3}\rangle\right)\,,\\
\label{4f}
&|\sigma_{3}\rangle = \frac{1}{\sqrt{3}}|2s\rangle -
\sqrt{\frac{2}{3}}\left(\frac{\sqrt{3}}{2}|2p_{x}\rangle +
|\chi_{3}\rangle\right)\,,\\
&|\chi_{3}\rangle=\frac{1}{2}|2p_{y}\rangle +\frac{\sqrt{3}a}{8R}|2p_{z}\rangle\,.
\end{eqnarray}
The $\pi$ orbitals are the same for the zig-zag and armchair nanotubes in
the lowest order of $a/R$. They are used to create the Bloch function
in the tight-binding approximation. As a result, we obtain  the following
$\pi$ orbital energy of the curved  graphene surface of radius $R$
\begin{eqnarray}
&&\varepsilon_{\pi} = \langle \pi|\verb"H"|\pi \rangle =\langle
2p_{z}|\verb"H"|2p_{z} \rangle +\frac{1}{24}\left(
\frac{a}{R}\right)^{2}\langle 2s|\verb"H"|2s \rangle\nonumber\\
&& + \frac{1}{48}\left( \frac{a}{R}\right)^{2}\langle
2p_{y}|\verb"H"|2p_{y} \rangle
= \varepsilon_{2p}+ \alpha
\left(\frac{a}{R}\right)^{2}\,,\\
&&\alpha  = \frac{1}{24}\langle s|\verb"H"|s \rangle +
\frac{1}{48}\langle p_{y}|\verb"H"|p_{y} \rangle\,.
\label{2}
\end{eqnarray}
Note, that the orbitals $2p_{y,z}, 2s$ are localized on the
same carbon atom and contribute to the $\pi$ orbital energy \cite{SDD},
while there is no such a contribution from the nondiagonal matrix
elements.
As a result, we obtain that the energy of the curved graphene consists of
the energy of the flat graphene $\varepsilon_{2p}$, and  the energy of the
$2s$, $2p_y$ orbitals brought about by the curvature.

Using the numerical values for the energies of the $|s\rangle$ and $|p_{y}\rangle$
orbitals  of the carbon atom
$\langle s|\verb"H"|s \rangle =-12$eV, $\langle p_{y}|\verb"H"|p_{y}
\rangle =-4$eV
(e.g., \cite{Lomer}),
we obtain for the parameter $\alpha \simeq -0.58$eV.

\end{document}